# Random Telegraph Noise of MIS and MIOS Silicon Nitride memristors at different resistance states

N. Vasileiadis, P. Loukas, A. Mavropoulis, P. Normand, I. Karafyllidis, G. Ch. Sirakoulis

and P. Dimitrakis

*Abstract*— Resistive memories (RRAM) are promising candidates for replacing present nonvolatile memories and realizing storage class memories; hence resistance switching devices are of particular interest. These devices are typically memristive, with a large number of discrete resistance levels and in some configurations, they perform as analog memories. Noise studies have been used to enlighten further the resistance switching mechanism in these devices. Recently, a strong interest arose on the noise generated by RRAM/memristors because due to their inherent stochasticity can be used for cryptography and other security applications. In this work, two fully CMOS compatible memristive devices both with silicon nitride as switching material, were examined in terms of multi-level resistance operation. Using appropriate SET/RESET pulse sequences through a flexible analog tuning protocol, four stable resistance states were investigated through random telegraph noise measurements and analysis. A possible model enlightening the role of silicon nitride defects in the observed noise signals at different resistive states is presented.

## I. INTRODUCTION

Resistive memories (RRAMs) are one of the most promising nonvolatile memory alternatives. RRAMs are appropriate for implementation in crossbar architecture, which is the most promising scenario to attain the lowest memory cell, $4F^2$ (F: the minimal feature size produced by lithography) [1] due to their outstanding scalability and simple two-terminal topology. It has been established that RRAM crossbar (Xbar) arrays can be used to create in-memory [2] and neuromorphic computing hardware accelerators [3]. We recently demonstrated that resistive switching (RS) devices can be employed as memristors in quantum simulators to store qubits [4]. Silicon nitride-based dielectrics are particularly attractive because of their resistance to humidity and oxygen-related parasitic effects, as well as metal ion diffusion [5 – 7]. In NVM technology, silicon nitride is a well-known material, and there is a variety of charge-trapping memory devices on the market (e.g., SONOS, BiCS). The presence of native bulk defects, which operate as charge-storage levels for both electrons and holes, explains its use in NVMs [6]. Silicon nitride has been recently established successfully for RRAM and memristors devices [7]. In addition, intentional doping inclusion can be used to adjust the characteristics of silicon nitride memristors

[6]. The development of a reliable accessing of numerous resistance levels is of primary importance for memory and computing applications. Furthermore, the fluctuation of any resistance level should be identified, and an accurate theory could predict its behavior. Recently it has been demonstrated that the noise signals generated by memristors either at LRS or HRS, can be successfully utilized for the implementation of hardware true-random number generators (TRNG) and physical unclonable functions (PUF) [8]. Thus, the investigations of noise signals in resistive switching memristors have been intensified the last years [9].

In this contribution, the role of the traps in resistive switching of MIS and MIOS memristive devices is discussed. The devices were firstly evaluated through I-V measurements and then Random Telegraph Noise (RTN) measurements and following, analysis was performed for different LRS. Finally, a charge trapping model is discussed to explain the observed RTN.

## II. DEVICE FABRICATION AND MEASUREMENT SETUP

$N^{++}$-type ($\rho < 0.005$ $\Omega \cdot$cm) 100mm (100) silicon wafers were used for MIS and MIOS device structures. The Si substrates were initially treated by the RCA "HF-last" cleaning process. On one $n^{++}$-Si wafer, a 2 nm thick tunnel barrier $SiO_2$ layer was thermally grown by dry oxidation in furnace at 850°C in O-diluted (10% $O_2$ v/v) ambient. Next, a 7 nm $Si_3N_4$ layer was deposited by LPCVD on both wafers at 810 °C. On another $n^{++}$-Si wafer, a 7nm $Si_3N_4$ layer was deposited by the same LPCVD procedure directly on the clean Si surface. Next, a 30 nm Cu / 30nm Pt top-electrode (TE) bilayer was deposited by sputtering on top of the nitride layers of MIOS ($Si_3N_4/SiO_2/n^{++}$-Si) and MIS ($Si_3N_4/n^{++}$-Si) wafers and patterned by a lift-off process. The TE square pads have dimensions of 100×100 μm$^2$. Pt was used to prevent the oxidation of the copper electrodes. A 500 nm thick Al metal layer was deposited on the back-side of the wafers, acting as bottom-electrode (BE). The $SiO_2$ layer, which was sandwiched between $SiN_x$ and Si introduces a greater energy barrier reducing the leakage of trapped carriers in bulk SiN to Si BE. Furthermore, the $SiO_2$/Si-BE interface has a better quality than the $SiN_x$/Si interface. As a result of thermal stimulation carrier exchange (trapping/de-trapping)

This work was supported in part by the research projects "3D-TOPOS" (MIS 5131411) and "LIMA-chip" (Proj.No. 2748) which are funded by the Operational Programme NSRF 2014-2020 and the Hellenic Foundation of Research and Innovation (HFRI) respectively.

Nikolaos Vasileiadis, Panagiotis Loukas, Alexandros Mavropoulis, Pascal Normand and Panagiotis Dimitrakis are with the Institute of Nanoscience and Nanotechnology, NCSR Demokritos, 15341 Agia Paraskevi, Attica, Greece (e-mail: n.vasiliadis, a.mavropoulis, p.normand, p.dimitrakis @inn.demokritos.gr).

Nikolaos Vasileiadis, Ioannis Karafyllidis and Georgios Ch. Sirakoulis are with the Electrical and Computer Engineering Department, Democritus University of Thrace, 67100 Xanthi, Greece (e-mail: nikvasil, ykar, gsirak @ee.duth.gr).

between BE and SiN$_x$ through boarder traps is minimized. It is envisaged that due to the presence of the SiO$_2$ layer, MIOS devices will exhibit less noisy electrical measurements compared to MIS. XPS analyses revealed that the SiN$_x$ material used in the examined devices has $x$=1.27 compared to 1.33 for the stoichiometric one.

The DC I-V characteristics of the fabricated devices were measured using the HP4155A and Tektronix 4200A. Impedance spectroscopy measurements in the range 100Hz – 1MHz were also performed on devices either in HRS or LRS using the HP4284A Precision LCR Meter. All measurements were performed at room temperature in ambient air (RH=45%). A special measurement setup and software was used to realize smart pulse tuning protocol, in order to achieve multi-resistance states as well as analog performance of the examined memristive devices, that is presented elsewhere [6]. RTN measurements were performed using a custom setup for recording the current under a constant read voltage, 0.1V, for 100s. A digital oscilloscope was used for current recording using a battery-powered SR570 I/V converter.

## III. RESULTS AND DISCUSSION

### A. Current-Voltage sweep measurements

Typical I-V sweep measurements for both types of devices are presented in Fig. 1. Evidently, all devices exhibited bipolar SET/RESET operation without any prior electroforming step. In addition, different resistance states can be achieved by varying the current compliance ($I_{CC}$) level. The goal of the $I_{CC}$ is to keep under control the current flowing through the nitride as approaching the TE – BE shorting (LRS) and thus avoid irreversible failure of the dielectric. During the RESET (negative voltage) sweep, the measured current exceeds the value of $I_{CC}$ used to SET the memristor (current overshoot). This is attributed to the measurement instrument's latency to stop the fast-current increase and to device's self-capacitance discharge [10]. Fig. 1(a) shows the I-V characteristics for MIS where current instabilities are observed during SET and RESET. Such instabilities are not presented in the corresponding I-V measurement on MIOS (Fig.1(b)), which should be attributed to the carrier exchange between nitride traps and BE. This is consistent with our initial considerations regarding the role of thin SiO$_2$ layer. The cumulative probability plots of SET/RESET voltages, $V_{SET}$/$V_{RST}$, for both device structures are presented in Fig.2. Obviously, $V_{SET}$ has better uniformity (smaller $\sigma/\mu$) compare to $V_{RST}$ in both devices, while MIS devices have slightly better uniformity compare to MIOS. Up to now, there is no definitive conclusion on the resistance switching mechanism in silicon nitride memristors.

The stoichiometry of the dielectric and the nature of the TE metal determine the switching mechanisms. According to the published research findings on silicon nitride memristors, the bipolar switching behavior was found independent on the nature of the TE [11]. Furthermore, bipolar switching occurs using either p$^{++}$-Si or n$^{++}$-Si as BE. The majority of the research articles conclude that the resistance switching mechanism proceeds from a trap-assisted mechanism [11] attributed to the traps existing in SiN$_x$.

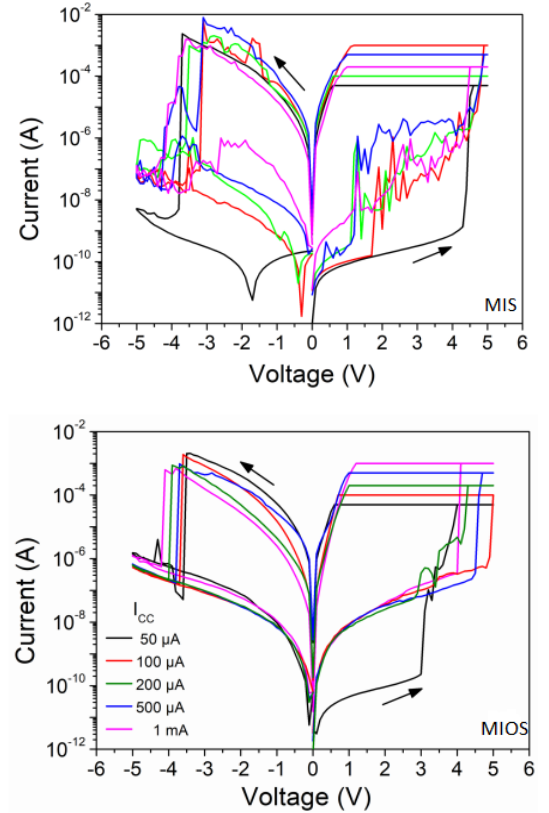

Figure 1. I-V sweep switching characteristics at different Icc for (a) MIS and (b) MIOS 1R cells.

The origin of these traps is due to deficiency of nitrogen, i.e., silicon dangling bonds and nitrogen vacancies $V_N$. During SET operation, defects are coordinated due to the electric field forming a conduction path (filament) for electrons injected from the n$^{++}$-Si BE [1]. More specifically, M. Yang et al. [11] concluded that N ions are attracted by the defect sites to the TE metal, forming a reservoir of N atoms, leaving behind Si dangling bonds. N ions are moving towards to the TE through the existing defects. At the end of the SET process a CF starting from the BE towards the TE is formed by the Si dangling bonds sites. According to their model, $V_{SET}$ and $V_{RST}$ depend on the reactivity of the TE metal with nitrogen. It is also well known that SiN is a diffusion blocker material for Cu [6]. Cu cations diffusivity in SiN is significantly lower than in SiO$_x$ [6]. DFT calculations revealed that the hopping energy required for a Cu cation to move from one site to another is much higher in SiN compared to SiOx [6]. Finally, the stoichiometry of the silicon nitride layer significantly affects the dielectric permittivity and the energy barrier with Si [6], as well as the concentration of traps [6].

### B. Noise Measurements

Using the flexible tuning method described in [5-7], the MIS and MIOS memristive devices were SET at four different resistance states, namely 60 kΩ, 200 kΩ, 500 kΩ and 800 kΩ. Then, the RTN signal was measured for each state using the experimental conditions described in Section II.

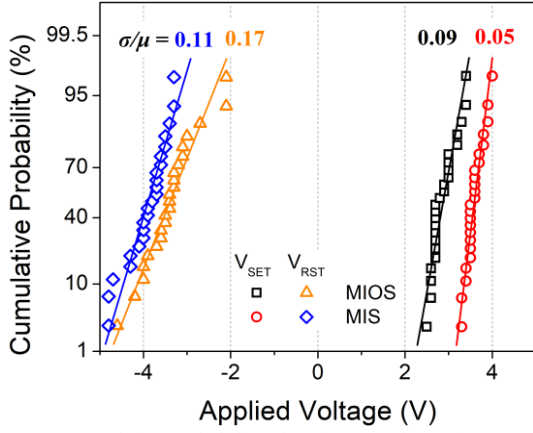

Figure 2. Cumulative probability plot of $V_{SET}$ and $V_{RST}$ for MIS and MIOS devices. Sample: 30 devices, $I_{CC}=100\mu A$

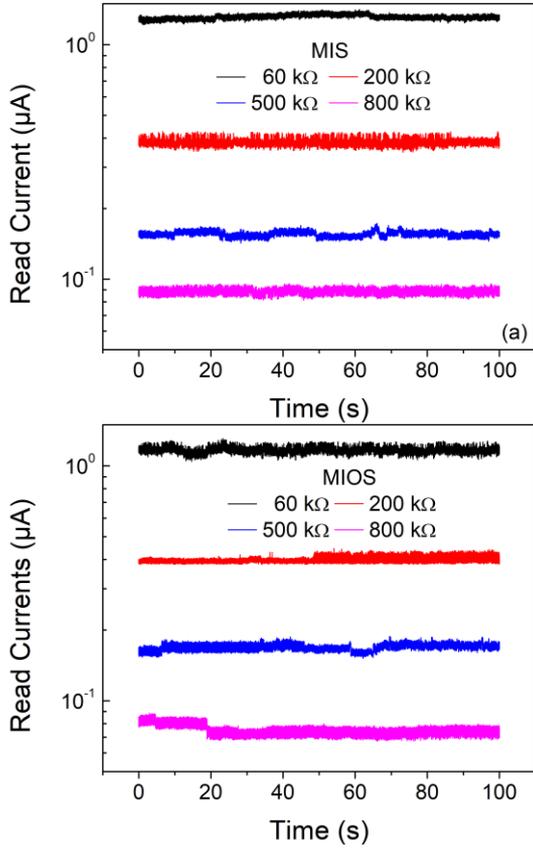

Figure 3. Current-time fluctuations measured at constant bias 0.1V for different resistance levels obtained for (a) MIS and (b) MIOS 1R cells.

The current fluctuation signals, in the time domain, are shown in Fig.3. Typical RTN spectra are presented in Fig.4. The signals were processed and analyzed by a custom Python program. Fig. 5 presents the power spectral density (PSD) $S_I$ for each individual recorded signal.

We may reasonably assume that the observed RTN signal in our devices mainly originates from the electron trapping – de-trapping process at defects surrounding the CF. The defects interact with electrons flowing in the CF as soon as they are inside their capture cross-section area, i.e., at a distance about one Debye length [9]. In order to study and analyze further the RTN observed at different resistance states in our filamentary type memrsitive devices, we compared the conductive filament (CF) formed between the top and bottom electrodes with a MOSFET.

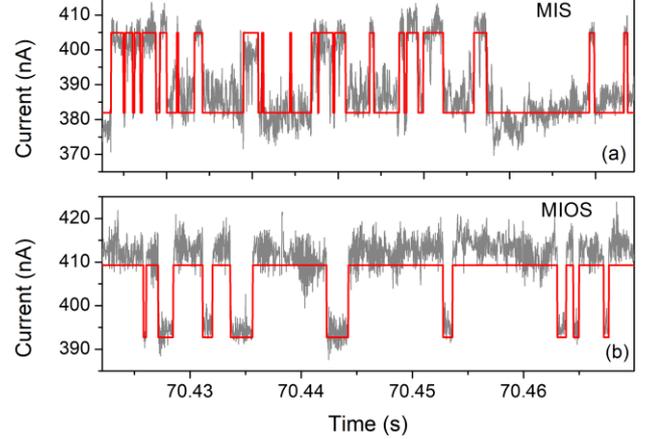

Figure 4. RTN signal recorded for MIS and MIOS samples corresponding to 200 kΩ resistance state, under constant bias 0.1V.

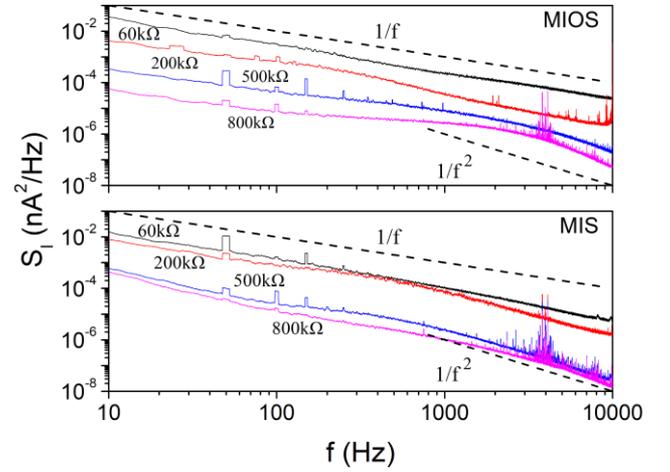

Figure 5. Power spectral density plots of the signals shown in Fig.3 for MIS and MIOS 1R cell.

TABLE I. RTN SIGNAL ANALYSIS

| RTN Signal Parameter | Resistance level (kΩ) | | | |
|---|---|---|---|---|
| | *60* | *200* | *500* | *800* |
| $\gamma$ (<3kHz) | 1.17[a]/1.00[b] | 0.73 / 0.89 | 0.80 / 0.75 | 0.99/ 0.42 |
| $\gamma$ (>3kHz) | 1.41 / 1.00 | 1.76 / 1.30 | 1.73 / 2.50 | 2.41/2.82 |
| State 1 (eV) | 0,68 / 0.63 | 0.60 / 0.63 | 0.68 / 0.67 | 0.64 / - |
| State 2 (eV) | 0.71 / 0,68 | 0.67 / 0.60 | 0.67 / 0.65 | 0.66 / - |

a. MIS b. MIOS

Obviously, measuring the noise of the CF in the bulk of SiN is equivalent to the noise measurement in a gate-all-around (GAA) nanowire MOSFET. In this context, $S_I$ corresponds to $S_{Id}$ and $V_{ds}$ corresponds to the applied read voltage (0.1V). The $V_{gs}$ modifies the channel conductance corresponding to the resistance of the filament. Evidently, the known noise theory of the MOSFET can be used to analyze the noise of the memristive devices and hence the $S_I$ of the observed RTN signals is defined by the equation:

$$S_I(f) = \langle \Delta I^2 \rangle \frac{4/f_T}{1+(f/f_T)^\gamma} \quad (1)$$

where $f_T$ is the relaxation frequency of the process and the exponent $\gamma$ is in the range $0 \leq \gamma \leq 2$ [12]. In case $\gamma=0$, the measured noise is a white noise. In case $\gamma=1$, the measured noise behaves as $1/f$ flicker noise. When $\gamma=2$, Eq. (1) has a Lorentzian form corresponding to the contribution of one strong trapping level to the RTN signal and $f_T$ is the frequency constant associated with the specific trapping level [13]. McWhorter *et al*. [14] demonstrated that $\gamma=1$ could be when there is a large number of different trapping levels, making the contribution of any individual level to the measured noise signal impossible to discern, i.e., $S_I(f) = \sum S_i(f) \sim 1/f$ where $S_i(f)$ is the PSD of the $i_{th}$ trapping level.

Following the analysis of RTN signals and related PSD curves, the trap time constant and the exponent $\gamma$ are estimated. The results are presented in Table I. For the case of 800 kΩ (very close to HRS), the exponent $\gamma > 2$ which cannot be explained by the trapping – de-trapping model. So large values revealing that internal instrument's noise has been amplified [14]. Also, values of $\gamma < 1$ were observed, suggesting that there is a large concentration of trap with characteristic frequency higher and that their distribution is skewed closer to CF/SiN interface [15]. As shown in Fig.5., two distinct frequency zones exist. Specifically, $\gamma$ is near to 1 at low-frequency regime (<3 kHz), and $\gamma > 1.5$ at higher frequencies (>3 kHz). Moreover, $\gamma$ is very close to unity for 60 kΩ and 200 kΩ LRS while deviations are observed for higher resistance states 500 kΩ and 800 kΩ. Adopting the analysis in [16] we can estimate the energy range of the traps participating to RTN. As shown in Table I, these are sited in the bulk of SiN between 0.60eV and 0.71eV. Also, the same levels were found for both MIS and MIOS devices indicating the origin of RTN in both device structures is due to the SiN bulk traps.

Fig.6 summarizes the possible origin of RTN in MIS and MIOS devices. At very low resistance the current flowing through the filament is strong enough to provide a sufficient number of electrons to interact with the surrounding traps having different energy levels and capture cross-sections. Thus, the measured noise signals have $1/f$ dependence. At medium resistance values of the CF, i.e., 200 kΩ and 500 kΩ, the number of electrons flowing through traps in the CF is severely smaller and hence only limited trapping states can interact with them, producing a noise signal of $\sim 1/f^{1.5}$. Finally, as CF resistance is moving towards to HRS, i.e., 800kΩ, there are only a few electrons available to interact with traps and consequently the noise signal has very poor density.

## IV. CONCLUSION

The resistance switching mechanism and the origin of RTN signal at different LRS in MIS and MIOS silicon nitride memristive devices is highlighted in this article. The role of traps and their contribution to RTN was discussed and accentuated. Finally, a model was formed to explain the role of traps in RTN.

ACKNOWLEDGMENT

This work was supported in part by the research projects "3D-TOPOS" (MIS 5131411) and "LIMA-chip" (Proj.No. 2748) which are funded by the Operational Program NSRF 2014-2020 and the Hellenic Foundation of Research and Innovation (HFRI) respectively.

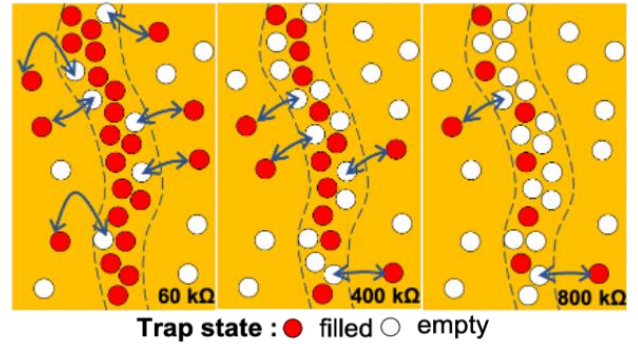

Figure 6. Schematic representation of the RTN mechanism due to different conductive filament resistance.